\documentclass[aps,prc,twocolumn,floatfix,nofootinbib,preprintnumbers,superscriptaddress,longbibliography]{revtex4-1}

\usepackage[T1]{fontenc}
\usepackage{latexsym}
\usepackage{amssymb}
\usepackage{amsfonts}
\usepackage{amsmath}
\usepackage{amsthm}
\usepackage{multirow}
\usepackage{longtable}
\usepackage{mathtools}
\usepackage{color}
\usepackage{ulem}
\usepackage{hyperref}
\usepackage{bm}
\usepackage[utf8]{inputenc}

\renewcommand{\vec}{\boldsymbol}

\makeatletter
\def\@bibdataout@aps{%
\immediate\write\@bibdataout{%
@CONTROL{%
apsrev41Control%
\longbibliography@sw{%
    ,author="08",editor="1",pages="1",title="0",year="1"%
    }{%
    ,author="08",editor="1",pages="1",title="",year="1"%
    }%
  }%
}%
\if@filesw \immediate \write \@auxout {\string \citation {apsrev41Control}}\fi 
}
\makeatother

\begin{document}

\title{$\beta^-{\rm p}$ and $\beta^-\alpha$ decay of the $^{11}$Be neutron halo ground state}

\author{J. Oko{\l}owicz}
\affiliation{Institute of Nuclear Physics, Polish Academy of Sciences, Radzikowskiego 152, PL-31342 Krak{\'o}w, Poland}

\author{M. P{\l}oszajczak}
\affiliation{Grand Acc\'el\'erateur National d'Ions Lourds (GANIL), CEA/DSM - CNRS/IN2P3, BP 55027, F-14076 Caen Cedex, France}

\author{W. Nazarewicz}
\affiliation{Department of Physics and Astronomy and FRIB Laboratory,
Michigan State University, East Lansing, Michigan  48824, USA}

\begin{abstract}
Beta-delayed proton emission from the neutron halo ground state of $^{11}$Be raised much attention due to the unusually high decay rate. It was argued that this may be due to the existence of a  resonance just above the proton decay threshold. In this Letter, we use the
 lenses of real-energy continuum shell model to describe several observables including the Gamow-Teller rates for the $\beta^-$-delayed $\alpha$  and proton decays, and argue that, within our model, the large $\beta^-{\rm p}$
branching ratio cannot be reconciled with other data.

\end{abstract}

\maketitle

\textit{Introduction--}
A very unusual decay, a $\beta^-$-delayed proton decay  of a {\it neutron-rich}  nucleus $^{11}$Be, predicted theoretically in \cite{Baye2011},  was studied in Refs.~\cite{Riisager2014,Ayyad2019,Riisager2020}. Experimentally,  the strength of this decay mode turned out to be 1.3(3)$\times 10^{-5}$ \cite{Ayyad2019}, i.e., several orders of magnitude higher than  predicted \cite{Baye2011}. This puzzle was explained \cite{Riisager2014} by the presence of a narrow resonance in $^{11}$B, recently found in Ref.~\cite{Ayyad2019} slightly above the proton separation energy $S_{\rm p}=11.2286$ MeV \cite{NNDC}. As estimated in Ref.~\cite{Ayyad2019}, in order to account for  the observed proton decay rate, this resonance must have a sizable single-proton content. In Ref.~\cite{Ayyad2019}, the resonance $J^{\pi}=(1/2^+,3/2^+$) was reported at $E=11.425(20)$\,MeV, i.e., 197(20)\,keV above the  one-proton emission threshold and only 29(20)\,keV below the one-neutron emission threshold. A recent study of Ref.~\cite{Riisager2020} suggests that the branching ratio for the  beta-delayed proton decay should be smaller than 2.2$\times 10^{-6}$, well below that of
\cite{Ayyad2019}, which adds to the puzzle.
Theoretically, the beta-decay rate estimate obtained in effective field theory~\cite{Elkamhawy2019} supports the  $\beta^-{\rm p}$ decay scenario while the shell model study of Ref.~\cite{Volya2020}
concludes that the reported half-life for the beta-delayed proton decay  is virtually impossible to explain. 

Recently, we have discussed the formation of a near-threshold $1/2^+_3$ resonance in $^{11}$B  \cite{Okolowicz2020} using the shell model  embedded in the continuum (SMEC), which is the open quantum system realization of the nuclear shell model (SM). Therein, we have argued that the presence of a proton resonance  in $^{11}$B, which happens to be  `conveniently' located near the proton threshold is not entirely unexpected; its existence  is the manifestation of quantum openness of the atomic nucleus. 
In contrast to a nearby $3/2^+_3$ resonance, the $1/2^+_3$ resonance carries a large imprint of the proton decay channel \cite{Okolowicz2020}.  In a recent R-matrix analysis of the $\beta^-$-delayed $\alpha$ particle decay of $^{11}$Be \cite{Refsgaard2019}, it was found that the observed $b_{\rm r}(\beta^-\alpha)$ branching ratio could not be explained without a contribution from the $3/2^+_3$ state. Thus the controversy about the rate of the $\beta^-{\rm p}$ decay cannot be resolved if the  $\beta^-\alpha$ decay branch is not considered \cite{Refsgaard2019}.  

\begin{figure}[htb]
\includegraphics[width=0.95\linewidth]{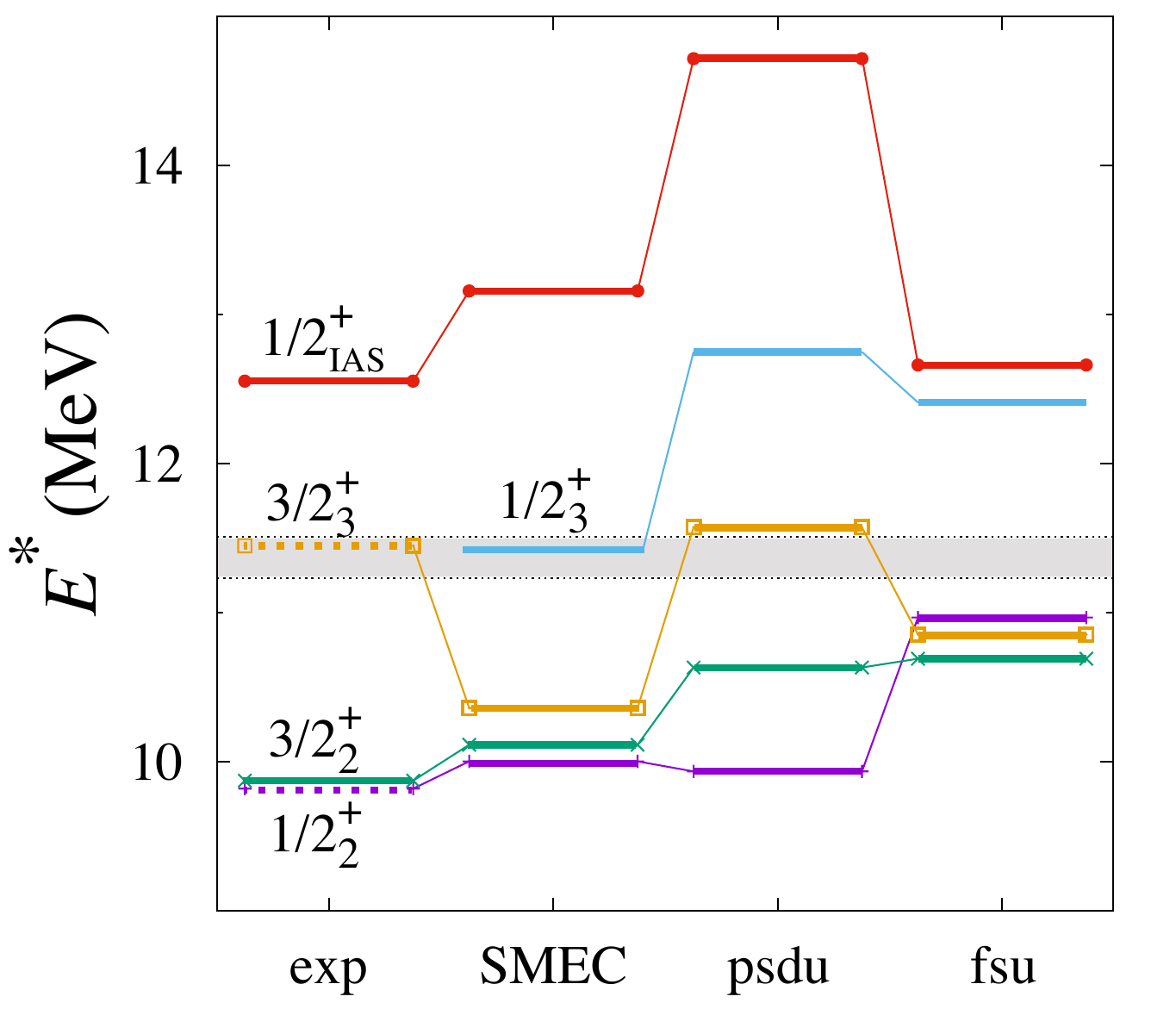}
\caption{Selected states in $^{11}\text{B}$ obtained with  SMEC and with psdu \cite{psdu} and fsu \cite{fsu} SM interactions used in Ref.\,\cite{Volya2020}. The  continuum coupling of SMEC  is $V_0 =  -37$\,MeV\,fm$^3$ with the energy scale adjusted to the energy of $1/2^+_3$ state taken from \cite{Ayyad2019}.
The shaded area represents the $\beta^-p$ window between  one-proton separation energy  and the  $Q$-value for $^{11}\text{Be}$ $\beta$-decay. The  assignments of experimental 1/2$^+_2$ and 3/2$^+_3$ states  are tentative/uncertain \cite{NNDC}; these levels are marked by a dotted line.
}
\label{exc}
\end{figure}

As seen in Fig.~\ref{exc}, in the accessible energy interval for the $\beta^-p$ decay, there is one resonance of energy $E=11.45(17)$ MeV, $\Gamma=93(17)$ keV with $J^{\pi}=(1/2^+,3/2^+)$  \cite{NNDC}. The key question is whether in this energy window there is only a single resonance with an angular momentum either $1/2^+$ or $3/2^+$, or there are two close-lying resonances with $J^{\pi}= 1/2^+$ and $J^{\pi}=3/2^+$. %

 The hypothesis of a single  $J^{\pi}=3/2^+$  resonance contradicts the observation of the $\beta^-$-delayed proton emission since the proton partial width in this case would be too small to enable detection of protons \cite{Ayyad2019}.  On the other hand, the hypothesis of a single $J^{\pi} = 1/2^+$ resonance contradicts the analysis of the $\beta^-$-delayed $\alpha$ emission \cite{Refsgaard2019}. Hence, the plausible way out of this dilemma is to assume the existence of two close-lying  resonances $1/2^+$ and $3/2^+$, decaying predominantly by  proton and $\alpha$ emission, respectively. This hypothesis does not contradicts the analysis of $\beta^-\alpha$ decay \cite{Refsgaard2019} and the observation of the $\beta^-{\rm p}$ decay \cite{Ayyad2019}. It is also consistent with the SMEC analysis \cite{Okolowicz2020}.
Hence, the controversy concerning the reported rate of $\beta^-$-delayed proton decay boils down to the consistency between 
the reported $b_{\rm r}(\beta^-{\rm p})$ and $b_{\rm r}(\beta^-\alpha)$ branching ratios. 

In this work, we test the scenarios put forward in Refs.~\cite{Ayyad2019,Refsgaard2019,Riisager2020}   by analyzing several observables in $^{11}$B, to constrain the value of the $\beta^-{\rm p}$ branching ratio for the decay of the halo ground state (g.s.)  of  $^{11}$Be. The effective SM interaction used has been adjusted to reproduce the experimental value of the branching ratio for $\beta^-$-delayed $\alpha$ decay \cite{Refsgaard2019}. The continuum-coupling strength has been adjusted assuming that the reported width $\Gamma = 12(5)$\,keV of the $J^{\pi} = 1/2^+$ resonance is equal to the proton-decay width. The  effective SMEC Hamiltonian determined in this way has been used to calculate the $\beta^-{\rm p}$ branching ratio and electromagnetic transitions in $^{11}$B. 

\vskip 0.2truecm
\noindent
\textit{SMEC--}
In the simplest version of SMEC, the Hilbert space is divided into two orthogonal subspaces ${\cal Q}_{0}$ and 
${\cal Q}_{1}$ containing 0 and 1 particle in the scattering continuum, respectively. An open quantum system  description of 
${\cal Q}_0$  includes couplings to the environment of decay channels through the energy-dependent effective Hamiltonian:
\begin{equation}
{\cal H}(E)=H_{{\cal Q}_0{\cal Q}_0}+W_{{\cal Q}_0{\cal Q}_0}(E),
\label{eq21}
\end{equation}
where $H_{{\cal Q}_0{\cal Q}_0}$ denotes the standard shell-model Hamiltonian describing the internal dynamics in the closed quantum system approximation, and 
\begin{equation}
W_{{\cal Q}_0{\cal Q}_0}(E)=H_{{\cal Q}_0{\cal Q}_1}G_{{\cal Q}_1}^{(+)}(E)H_{{\cal Q}_1{\cal Q}_0},
\label{eqop4}
\end{equation}
is the energy-dependent continuum coupling term, where $E$ is a scattering energy, $G_{{\cal Q}_1}^{(+)}(E)$ is the one-nucleon Green's function, and ${H}_{{Q}_0,{Q}_1}$ and ${H}_{{Q}_1{Q}_0}$ couple ${\cal Q}_{0}$ with ${\cal Q}_{1}$. 
The energy scale in (\ref{eq21}) is defined by the lowest one-nucleon emission threshold. The channel state is defined by the coupling of one nucleon in the scattering continuum to a shell model wave function of the nucleus  $(A-1)$. 

\vskip 0.2truecm
\noindent
\textit{Interaction--}
For the SM Hamiltonian $H_{{\cal Q}_0{\cal Q}_0}$ we adopted the  WBP$-$  interaction in the full $psd$ model space \cite{Yuan2017} with the  inert core of $^4$He. As discussed below,  the  monopole terms ${\cal M}^{\rm T=1}(0p_{1/2}1s_{1/2})$ and ${\cal M}^{\rm T=1}(0p_{1/2}0d_{5/2})$ have been adjusted to reproduce the experimental branching ratio 
 $b_{\rm r}(\beta^-\alpha)$.

The continuum-coupling interaction was assumed to be the  Wigner-Bartlett contact force
$V_{12}=V_0 \left[ \alpha + \beta P^{\sigma}_{12} \right] \delta\langle\vec{r}_1-\vec{r}_2\rangle$, where $\alpha + \beta = 1$ and $P^{\sigma}_{12}$ is the spin exchange operator. 
The value of $V_0$ was constrained by the measured proton width of the  $1/2^+_3$ state.
The spin-exchange parameter $\alpha$ has a value $\alpha = 2$ which is appropriate for nuclei in the vicinity of driplines \cite{Luo2002,Michel2004,Charity2018}.

The radial single-particle wave functions (in ${\cal Q}_0$) and the scattering wave functions (in ${\cal Q}_1$) are generated by the Woods-Saxon potential, which includes spin-orbit and Coulomb terms. The radius and diffuseness of the Woods-Saxon potential are $R_0=1.27 A^{1/3}$\,fm and $a=0.67$\,fm, respectively. The strength of the spin-orbit potential is $V_{\rm SO}=7.62$\,MeV,  and the Coulomb term is that of a uniformly charged sphere with radius $R_0$. The depth of the central potential  is adjusted to reproduce the measured separation energies of the $1s_{1/2}$ and $0d_{3/2}$ orbits. 
\vskip 0.2truecm
\noindent
\textit{Excited states of  $^{11}${\rm Be} and $^{11}${\rm B}--}Table~\ref{tab1}
shows the lowest excited states in $^{11}$Be calculated in SMEC.  The predicted decay width  of $5/2_1^+$ resonance is calculated at the measured energy.  Considering the fact that  $^{11}$Be is a halo system in its $1/2^+$ ground state that  requires huge configuration space in SM \cite{Calci2016},  the agreement between SMEC and experiment is satisfactory.
\begin{table}[htp]
\caption{Energy  and neutron decay width (both in keV) of  the excited states $1/2^-_1$ and $5/2^+_1$ in $^{11}$Be calculated in SMEC and compared to experiment. The theoretical error bars are due to the uncertainty in the determination of the continuum-coupling strength $V_0$, see below. }
\begin{ruledtabular}
\begin{tabular}{ccccc}
$J^{\pi}$~~ & $E^{(\rm th)}$  & $E^{(\rm exp)}$ & ${\Gamma}^{(\rm th)}$  & ${\Gamma}^{(\rm exp)}$      \cr
\hline
$1/2^-_1$ & ~$583^{+8}_{-10}$ & ~~320.04(10) & --  & --     \cr
$5/2^+_1$ & $2172^{+13}_{-20}$ & 1783(4)  & ~$98^{-51}_{+80}$ &~100(10)     \cr
\end{tabular}
\end{ruledtabular}
\label{tab1}
\end{table}%

Figure~\ref{exc} shows experimental and calculated (our SMEC model and the psdu and fsu models  of  Ref.\,\cite{Volya2020}) states of $^{11}$B around the  $\beta^-{\rm p}$  window of $^{11}\text{Be}$.
Currently, no {\it ab initio} calculations exist for these positive-parity  resonances in $^{11}$B to compare with the results of models shown in Fig.~\ref{exc}, see, e.g.,  the recent papers \cite{Choudhary2020,Soma2020}. Our calculations predict two 1/2$^+$ and two 3/2$^+$ excited states  that can be fed by the $\beta^-$  decay of $^{11}$Be. The  $T=3/2$ isobaric analog state (IAS)  \cite{Fortune2012} is calculated to lie over 1\,MeV above  the  $\beta^-p$  window in all models.
It is seen that both the predicted  energies and ordering  of these states  strongly depend on the specific interaction used. Consequently, due to strong  energy dependence of $\beta$- and electromagnetic decay rates, in the following discussion the  energy relations are taken from experiment while wave functions -- from SMEC. In this respect, we closely follow the procedure used  in Ref.\,\cite{Volya2020}. We wish to emphasize that -- albeit fairly phenomenological --  the SMEC is at present the only configuration interaction  approach that is capable of  describing the coupling to the scattering continuum and providing an accurate description of positive-parity resonances  in $^{11}$B.

\begin{figure}[htb]
\includegraphics[width=1.0\linewidth]{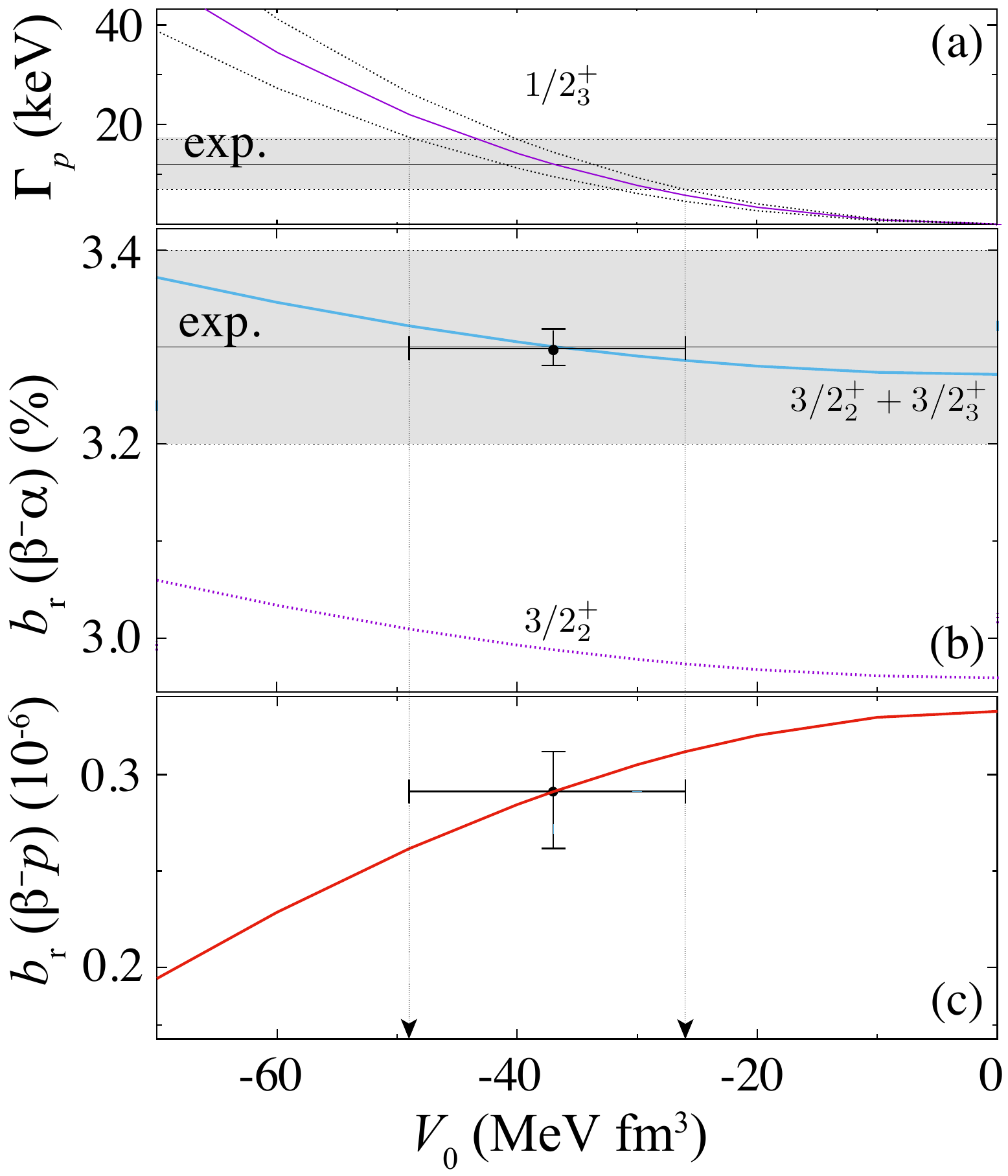}
\caption{(a) Proton decay width of the  $1/2^+_3$  resonance in $^{11}$B calculated in SMEC as a function of $V_0$ at an experimental energy $E = 11425(20)$\,keV. The experimental value of $\Gamma(1/2^+_3)$ \cite{Ayyad2019}   is marked by horizontal lines. The vertical arrows mark the range of $V_0 =  -37^{+11}_{-12}$ MeV fm$^3$  constrained by the measured value of $\Gamma_{\rm p}$. The uncertainty of the calculated proton width due to the experimental error on the resonance energy is shown by thin dotted curves. (b) The branching ratio for the $\beta^-$-delayed $\alpha$ emission from the $1/2^+$  g.s.\ of $^{11}$Be  to the $3/2_2^+$, and $3/2_3^+$ resonances in $^{11}$B. The predicted value of $b_{\rm r}(\beta^-\alpha)$ is constrained by the proton emission width of $1/2^+_3$ resonance.  The experimental value of  $b_{\rm r}(\beta^-\alpha)$ \cite{Refsgaard2019}  is 3.30(10) $\%$. (c) The branching ratio $b_{\rm r}(\beta^-p)$.  The energy dependence of the  resonance width is given by 
Eq.~(\ref{ourgamma}).
}
\label{fig1}
\end{figure}

\vskip 0.2truecm
\noindent
\textit{Proton decay width of the $1/2_3^+$ resonance in $^{11}${\rm B}--}As discussed in \cite{Okolowicz2020} the $1/2^+_3$ resonance is interpreted as a state aligned to the proton-decay channel. Consequently, its total decay width is expected to be dominated by the proton partial width, i.e., $\Gamma\approx \Gamma_{\rm p}$. In SMEC
$\Gamma_{\rm p}$  of this state depends weakly on the variations of monopole terms ${\cal M}^{\rm T=1}(0p_{1/2}1s_{1/2})$ and ${\cal M}^{\rm T=1}(0p_{1/2}0d_{5/2})$ but rapidly changes with $V_0$. This allows to use the experimental value of $\Gamma(1/2^+_3) = 12(5)$\,keV \cite{Ayyad2019} to constrain the continuum coupling strength. As see in
Fig.~\ref{fig1}(a), this value  is $V_0 =  -37^{+11}_{-12}$ MeV fm$^3$; it is used in all calculations in this work.  
 
\vskip 0.2truecm
\noindent
\textit{Branching ratio for the $\beta^- \alpha$ decay of  $^{11}${\rm Be}--}
The range of energies in $^{11}$B accessible in the $\beta^-\alpha$  decay of $^{11}$Be 
 encompasses the range of energies accessible in the $\beta^-{\rm p}$ decay. 
In the energy window for the $\beta^-\alpha$ decay, one can find $3/2^+_2$ and $3/2^+_3$  resonance, $(1/2^+_2)$ resonance lying close to the  $3/2^+_2$ state \cite{NNDC}, and $1/2^+_3$ resonance 
reported in Ref.~\cite{Ayyad2019}. As seen in Fig.~\ref{exc}, a doublet of close-lying states $1/2^+_2$, $3/2^+_2$ is also predicted by our SMEC calculations.

The $\beta^-$-delayed proton decay is a two-step process where $^{11}$Be in the ground state decays to $^{10}$Be via an intermediate resonance in $^{11}$B. In our branching-ratio estimates we only consider Gamow-Teller  transitions. 
As will be shown below, the Fermi component of the beta decay via the tail of the  1/2$^+$, $T=3/2$  IAS can be safely neglected. 
 Branching ratios are calculated as:
 \begin{equation}
 b_{\text r} = \frac{{\text T}_{1/2}}{\cal T} \,{\cal F}_0 \,g^2_A\, B({\text GT})
 \label{brform}
 \end{equation}
 where ${\text T}_{1/2} = 13.76$\,s is the half-life time of $\beta^-$-decaying ground state of $^{11}$Be, ${\cal T}=6145$\,s, $g_A = 1.27$, and 
  the phase space factor ${\cal F}_0$ for this sequential two-step process can be written as:
 \begin{equation}
 {\cal F}_0 = \int^Q_0 \frac{{\text d} \epsilon}{2 \pi}\, \frac{f_0(Q-\epsilon)\, \Gamma(\epsilon)}{(\epsilon - E_R)^2 + \Gamma^2(\epsilon)/4},
 \label{phasesp}
 \end{equation}
 where $f(\epsilon)$ is  the energy-dependent phase space function,  $E_R$ is the resonance's energy, $\Gamma$ is the decay width, and $Q$ is the maximum energy allowed for a given decay channel. The energy dependence of $\Gamma$  is essential
 for attenuating the function $f_0$   in the limit $\epsilon \rightarrow 0$.
 
The energy dependence of $\Gamma$ can be approximated by:
\begin{equation}
\Gamma(\epsilon) = \Gamma_R \frac{P_\ell (\epsilon)}{P_\ell (E_R)},
\label{gammae}
\end{equation}
where $\Gamma_R$ is an experimental width of the resonance, and $P_\ell$ is the barrier penetrability given by the standard $R$-matrix expression \cite{Lane1958}:
\begin{equation}
P_\ell (k) = \frac{k\,R_C}{F_\ell^2(k\,R_C)+G_\ell^2(k\,R_C)}
\label{penetr}
\end{equation}
where $F$ and $G$ are Coulomb regular and irregular wave functions, $\ell$ is the angular momentum of the emitted particle, and $R_C=1.57{\text A}^{1/3}$ is the  Coulomb radius~\cite{Wilkinson1962}. 
If the continuum space of SMEC contains the particle's decay channel, the energy dependence of resonance width is calculated using the imaginary part of the complex-energy solution ${\cal E}(E)$ of the SMEC effective Hamiltonian:
\begin{equation}
\Gamma(\epsilon) = -2 \Im[\,{\cal E}  (\epsilon)]. 
\label{ourgamma}
\end{equation} 
In our calculation of $b_{\rm r}(\beta^-\alpha)$, we use  Eq.~(\ref{phasesp}) to calculate the contribution of the $3/2^+_2$ and $3/2^+_3$ resonances. 
As the  discussed  $3/2^+$ resonances in $^{11}$B   $\alpha$-decay to low-energy negative parity states of $^7\text{Li}$, 
we take $\ell=1$  in Eq.~(\ref{penetr}). 

\begin{table}[htp]
\caption{Gamow--Teller reduced matrix elements $B(GT)$ and $\log_{10}(ft)$ values for the beta decay of g.s. of $^{11}$Be to the low-lying  states of $^{11}$B predicted in SMEC. The value $\log_{10}(ft)$ for the $T=3/2$ IAS contains the contribution from the Fermi transition. For comparison, the $B(GT)$ predictions  of Ref.~\cite{Volya2020} with the psdu SM Hamiltonian are also shown.}
\begin{ruledtabular}
\begin{tabular}{cccc}
{~$J^{\pi}$~} & $B(GT)$ \cite{Volya2020} & $B(GT)$ this work & $\log_{10}(ft)$ \\ \\[-8pt]
\hline 
$1/2^+_3$ & $0.659$ & $0.125^{+0.009}_{-0.013}$ & $4.485$  \\
$3/2^+_2$ & $0.001$ & $0.240^{-0.001}_{+0.002}$ & $4.200$ \\
$3/2^+_3$ & $0.615$ & $0.322^{+0.002}_{-0.003}$ & $4.073$ \\
\hline
$1/2^+,\,T=3/2$ & $0.571$ & $0.166^{+0.001}_{-0.001}$ & $3.274$ 
\end{tabular}
\end{ruledtabular}
\label{tab2x}
\end{table}%
 The $B(GT)$ matrix elements in SMEC are calculated between the g.s.\ of $^{11}$Be and the relevant $1/2^+$ and $3/2^+$ eigenstates of $^{11}$B shown in Table \ref{tab2x}. The SMEC g.s.\ of $^{11}$Be is obtained by coupling five lowest energy SM $1/2^+$ states to the lowest neutron decay channel. Similar procedure is followed for the excited $1/2^-_1$ and $5/2^+_1$ states of $^{11}$Be. The $1/2^+$ and $3/2^+$ SMEC eigenstates of $^{11}$B are obtained by mixing the lowest energy $1/2^+$ and $3/2^+$ SM states, respectively, via the coupling to the lowest one-proton and one-neutron decay channels.
The  coupling of SM states to the $\alpha$-decay channel cannot be described by the current realization of SMEC and in the calculation of the branching ratio 
$b_{\rm r}(\beta^-\alpha)$ we take experimental values of the resonance widths for $\Gamma_{\rm R}$ in (\ref{gammae}).
Table~\ref{tab2x} lists  the predicted $B(GT)$ values and  half-lives for the beta decay of g.s. of $^{11}$Be to the low-lying  states of $^{11}$B. 

Figure \ref{fig1}(b) shows the $V_0$-dependence of $b_{\rm r}(\beta^-\alpha)$ from the $1/2^+$    g.s.\ of $^{11}$Be to the $3/2^+_2$ and $3/2^+_3$ resonances in $^{11}$B. 
In agreement with results of the R-matrix analysis of $\beta$-delayed $\alpha$ spectrum \cite{Refsgaard2019}, only $3/2^+_2$ and $3/2^+_3$ resonances play a role in the $\beta^-\alpha$ decay of $^{11}$Be. The $1/2^+_3$ resonance does not add significantly to this branching ratio. Assuming that its partial $\alpha$-decay width $\Gamma_{\alpha}$ is 10 keV, the contribution of this resonance to $b_{\rm r}(\beta^-\alpha)$  is $1.7\cdot10^{-4}$ and can be safely neglected.
As mentioned earlier, in our calculations, we took the experimental value  of 9.873\,MeV for the energy of $3/2_2^+$ resonance \cite{NNDC}. For the $3/2_3^+$ state, following the suggestion of Refsgaard et al.\ \cite{Refsgaard2019}, we assumed the energy of $11.450$\,MeV. 

As seen in Fig.~\ref{fig1}(b), major contribution to $b_{\rm r}(\beta^-\alpha)$ comes from the $3/2^+_2$ state, but the  $3/2^+_3$ states contributes coherently. In the psdu calculations of Ref.~\cite{Volya2020}, the contribution of the $3/2^+$ resonances is  $0.054 \%$, and the $1/2^+_3$ resonance adds to the  branching ratio $9\cdot10^{-4}\,\%$ for $\Gamma_{\alpha}=10$\,keV. 
The calculated $b_{\text r}(\beta^-\alpha)$ branching ratio varies very slowly with $V_0$. For that reason $b_{\text r}(\beta^-\alpha)$ has been used to fine-tune the monopole terms  of the SM interaction. To obtain a
satisfactory agreement with experimental branching ratio, ${\cal M}^{\rm T=1}(0p_{1/2}1s_{1/2})$ and ${\cal M}^{\rm T=1}(0p_{1/2}0d_{5/2})$
were shifted by $-2.292$\,MeV and  $+1.0$\,MeV, respectively.

\vskip 0.2truecm
\noindent
\textit{Branching ratio for the $\beta^-{\rm p}$ decay of} $^{11}$Be--
 In Ref.~\ \cite{Ayyad2019}, the reported $\beta^-{\rm p}$ branching ratio is $1.3(3)\cdot 10^{-5}$. 
 Figure \ref{fig1}(c) shows this branching ratio calculated in SMEC. Contrary to  $b_{\rm r}(\beta^-\alpha)$,  $b_{\rm r}(\beta^-{\rm p})$ strongly depends  on $V_0$  but the dependence on $T=1$ monopole terms  is weak. 
As one can see,  the predicted value of $b_{\rm r}(\beta^-{\rm p})$ is by a factor of $\sim 40$ smaller than found in Ref.\,\cite{Ayyad2019}. 

\begin{figure}[htb]
\includegraphics[width=1.0\linewidth]{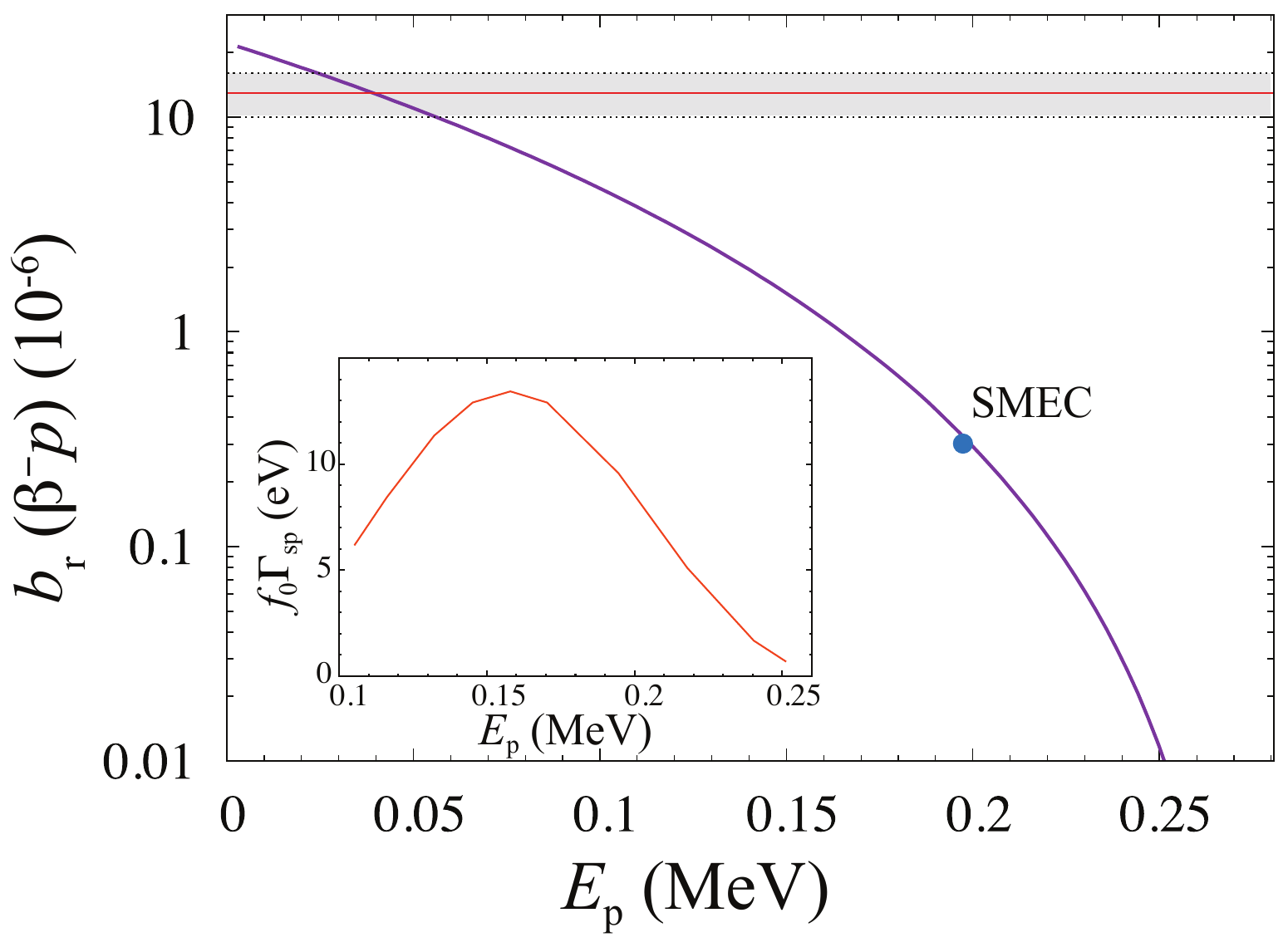}
\caption{Dependence of the branching ratio $b_{\rm r}(\beta^-{\rm p})$ calculated in  SM on the resonance energy $E_{\rm p}$ counted with respect to the proton emission threshold $E_{\rm p}=0$.
The SMEC prediction  at the reported  resonance energy \cite{Ayyad2019} is marked. Horizontal lines indicate experimental value of $b_{\rm r}(\beta^-{\rm p})$. The product $f_0\,\Gamma_{\text sp}$ is plotted in the insert.
}
\label{br3}
\end{figure}

In order to shed light on this discrepancy,  in Fig.~\ref{br3} we show the $b_{\rm r}(\beta^-{\rm p})$ branching ratio obtained in SM as a function of the  energy of the $1/2^+_3$ resonance in the energy window of the $b_{\rm r}(\beta^-{\rm p})$ decay. One can see that the reported value of $b_{\rm r}(\beta^-{\rm p})$ is reached at $E_p\sim50$ keV above the proton emission threshold. At this energy, the proton decay width of $1/2^+_3$ resonance is expected to  be significantly smaller than the value given in Ref.~\cite{Ayyad2019}, i.e., the proton decay width and the branching ratio $b_{\rm r}(\beta^-{\rm p})$ seem to  be mutually inconsistent.
To further illustrate the dependence of $b_{\rm r}(\beta^-{\rm p})$ on the energy of $1/2^+_3$ resonance, we show in the insert of Fig.~\ref{br3} the product $f_0\,\Gamma_{\rm sp}$, which reflects the interplay between  the  phase space of the $\beta$ decay and a proton tunneling probability. The energy dependence of the single-particle width $\Gamma_{\rm sp}(E_p)$ is calculated by changing the depth of the  proton Woods-Saxon potential.
The quantity  $f_0\,\Gamma_{\rm sp}$ suggests that the highest $b_{\rm r}(\beta^-{\rm p})$ should be seen at $E_{\rm p} \sim 160$\,keV, close to the value expected from the SMEC analysis~\cite{Okolowicz2020} and  not far from the value found in Ref.\,\cite{Ayyad2019}. Still, according to our model, the experimental values of $\Gamma_{\rm p}$ and $b_{\rm r}(\beta^-{\rm p})$ cannot be reconciled.

\vskip 0.2truecm
\noindent
\textit{Electromagnetic transitions in $^{11}${\rm B}--}
The SMEC effective Hamiltonian fine-tuned to the decay data in $^{11}$B  can be used to calculate the electromagnetic transitions involving the low-energy states of $^{11}$B. This is an independent test of the SMEC approach and the optimized interaction.

Table \ref{tab11} compares experimental and SMEC total $\gamma$-decay widths for several low-lying states of both parities.
\begin{table}
\caption{Theoretical radiative widths calculated in SMEC for low-energy states in $^{11}$B  compared to experimental data \cite{NNDC}. The theoretical errors due to the uncertainty of $V_0$  are negligible. }
\begin{ruledtabular}
\begin{tabular}{ccc}
~~~$J^\pi$~~~ & ~~~${\it \Gamma}_\gamma^{\mbox{\tiny (th)}}$ (eV) ~~~& ~~~${\it \Gamma}_\gamma^{\mbox{\tiny (exp)}}$ (eV) ~~~ \\
\hline
1/2$^-_1$ & 0.244 & $0.12\pm 0.004$   \cr 
5/2$^-_1$ & 0.358 & $0.55\pm 0.05$   \cr
3/2$^-_2$ & 1.494 & $1.97\pm 0.07$   \cr
7/2$^-_1$ & 0.025 & $0.03\pm 0.007$  \cr
1/2$^+_1$ & 0.101 & $0.39\pm 0.05$    \cr
5/2$^+_1$ & 1.111 & $1.14\pm 0.08$    \cr
3/2$^+_1$ & 0.201 & $1.15\pm 0.15$   \cr
\end{tabular}
\end{ruledtabular}
\label{tab11}
\end{table}
One can see that the overall agreement between experimental and calculated values is good.
One way to clarify the existence and nature of the near-threshold $1/2^+_3$ resonance would be through the observation of gamma transitions from this state to different proton-bound states in $^{11}$B. Table \ref{tab0} lists the SMEC partial $\gamma$-decay widths and intensities of the electromagnetic transitions depopulating the  $1/2^+_3$ resonance. 
\begin{table}[htp]
\caption{Radiative widths and intensities of the electromagnetic transitions depopulating the  $1/2^+_3$ threshold resonance in $^{11}$B calculated in SMEC  with $\alpha = 2$. The SM width is ${\it \Gamma}_\gamma^{\mbox{\tiny total}} = 13.98$\,eV.
}
\begin{ruledtabular}
\begin{tabular}{ccccc}
$J^\pi_{\mbox{\tiny final}}$ & $E_\gamma$ (MeV) & M($\gamma$) &  ${\it \Gamma}_\gamma^{\mbox{\tiny partial}}$ (eV) & $I(\gamma)$\,(\%) \\[+2pt]
\hline 
3/2$^-_1$ & 11.425 & E1 & $0.803$ & $5.6 $ \cr
1/2$^-_1$ & 9.300 & E1 & 11.7 & $81.7$ \cr
3/2$^-_2$ & 6.405 & E1 & 1.61 & $11.2$ \cr 
3/2$^-_3$ & 2.865 & E1 & $5.3\cdot 10^{-5}$ &  $3.7\cdot 10^{-4}$ \cr 
1/2$^+_1$ & 4.633 & M1 & $8.6\cdot 10^{-2}$ & $0.6$ \cr 
5/2$^+_1$ & 4.139 & E2 & $3.7\cdot 10^{-3}$ & $2.6\cdot 10^{-2}$ \cr 
3/2$^+_1$ & 3.447 & M1 & $0.112$ & 0.78 \cr
 &                                                  & E2 & $9.8\cdot 10^{-6}$ & $6.8\cdot 10^{-5}$ \cr 
5/2$^+_2$ & 2.153 & E2 & $6.0\cdot 10^{-7}$ & $4.2\cdot 10^{-6}$ \cr
\cline{3-4}
 & & \multicolumn{2}{c}{${\it \Gamma}_\gamma^{\mbox{\tiny total}} = 14.32^{-0.16}_{+0.25}$} & \\[+2pt]
\end{tabular}
\end{ruledtabular}
\label{tab0}
\end{table}
For these estimates, we took the reported energy of $1/2^+_3$ resonance and the experimental energies of the final states \cite{NNDC}. One can see that E1 transitions to $1/2^-_1$ and $3/2^-_2$  states are particularly enhanced; those are possible candidates to prove the existence of $1/2^+_3$  resonance.

\vskip 0.2truecm
\noindent
 \textit{Conclusions--} Different experimental information on $^{11}$B and $^{11}$Be allow to constrain parameters of the effective SMEC Hamiltonian and investigate their internal consistency in the description of the $\beta^-{\rm p}$ decay of $^{11}$Be. 
 The  assumption in our study is that there exists the resonance that lies  between proton and neutron decay thresholds~\cite{Ayyad2019,Okolowicz2020}. It is also assumed that the nearby $3/2^+_3$ resonance, which is seen in the $\beta^-$-delayed $\alpha$-particle emission \cite{Refsgaard2019}, couples weakly to the one-proton decay channel, and hence cannot significantly contribute to the $\beta^-{\rm p}$ decay of $^{11}$Be. 
 
 In this Letter, we have shown that  the data on branching ratio for $b_{\rm r}(\beta^-\alpha)$  \cite{Refsgaard2019} and  $\Gamma_p(1/2^+_3)$ \cite{Ayyad2019}  can be consistently described.  With the fine-tuned SMEC  parameters,  one obtains a reasonable description of both the electromagnetic transitions in $^{11}$B and the low-energy excitations in $^{11}$Be. However, the branching ratio $b_{\rm r}(\beta^-{\rm p})= 0.28(3)\times 10^{-6}$ calculated with this Hamiltonian disagrees with the reported value $1.3(3)\times 10^{-5}$ \cite{Ayyad2019}. We thus conclude that there is a tension between the
  reported values of $b_{\rm r}(\beta^-\alpha)$ and $\Gamma_p(1/2^+_3)$,  and with $b_{\rm r}(\beta^-{\rm p})$.
  
  The explanation of a strong $b_{\rm r}(\beta^-{\rm p})$ branch as a decay through the tail of an IAS  does not seem to be a plausible alternative: for the continuum coupling strength that is consistent with $\Gamma_{\rm p}(1/2^+_3)$, the partial proton decay width of the IAS obtained in SMEC is $\Gamma_{\rm p} \approx 26$\,keV, which is much too narrow to contribute significantly to $b_{\rm r}(\beta^-{\rm p})$. This conclusion agrees with findings of Ref.~\cite{Volya2020}.
 
In summary, an independent experimental investigation is called for that would confirm or reject the existence of the $1/2^+_3$ resonance and confirm the presence of the $3/2^+_3$ resonance. We hope that our theoretical estimates will stimulate such a search.

After submitting this Letter, we were happy to learn that
a near-threshold resonance in $^{11}{\rm B}$ at $E\approx 11.44$\,MeV was reported in the proton resonance scattering \cite{PhysRevLett.129.012501} and in   $^{10}{\rm Be}(d,n)\rightarrow ^{10}{\rm Be} + p$    \cite{PhysRevLett.129.012502} reactions. Both findings are consistent with our predictions.

\vskip 0.2truecm
\noindent
{\it Acknowledgements}---
We thank A. Volya for useful discussions. This material is based upon work supported by
the U.S. Department of Energy, Office of Science, Office of Nuclear Physics under Award No.DE-SC0013365 (Michigan
State University) and by the COPIN and COPIGAL French-Polish scientific exchange programs.

\bibliography{threshold}

\begin{thebibliography}{23}%
\makeatletter
\providecommand \@ifxundefined [1]{%
 \@ifx{#1\undefined}
}%
\providecommand \@ifnum [1]{%
 \ifnum #1\expandafter \@firstoftwo
 \else \expandafter \@secondoftwo
 \fi
}%
\providecommand \@ifx [1]{%
 \ifx #1\expandafter \@firstoftwo
 \else \expandafter \@secondoftwo
 \fi
}%
\providecommand \natexlab [1]{#1}%
\providecommand \enquote  [1]{``#1''}%
\providecommand \bibnamefont  [1]{#1}%
\providecommand \bibfnamefont [1]{#1}%
\providecommand \citenamefont [1]{#1}%
\providecommand \href@noop [0]{\@secondoftwo}%
\providecommand \href [0]{\begingroup \@sanitize@url \@href}%
\providecommand \@href[1]{\@@startlink{#1}\@@href}%
\providecommand \@@href[1]{\endgroup#1\@@endlink}%
\providecommand \@sanitize@url [0]{\catcode `\\12\catcode `\$12\catcode
  `\&12\catcode `\#12\catcode `\^12\catcode `\_12\catcode `\%12\relax}%
\providecommand \@@startlink[1]{}%
\providecommand \@@endlink[0]{}%
\providecommand \url  [0]{\begingroup\@sanitize@url \@url }%
\providecommand \@url [1]{\endgroup\@href {#1}{\urlprefix }}%
\providecommand \urlprefix  [0]{URL }%
\providecommand \Eprint [0]{\href }%
\providecommand \doibase [0]{http://dx.doi.org/}%
\providecommand \selectlanguage [0]{\@gobble}%
\providecommand \bibinfo  [0]{\@secondoftwo}%
\providecommand \bibfield  [0]{\@secondoftwo}%
\providecommand \translation [1]{[#1]}%
\providecommand \BibitemOpen [0]{}%
\providecommand \bibitemStop [0]{}%
\providecommand \bibitemNoStop [0]{.\EOS\space}%
\providecommand \EOS [0]{\spacefactor3000\relax}%
\providecommand \BibitemShut  [1]{\csname bibitem#1\endcsname}%
\let\auto@bib@innerbib\@empty
\bibitem [{\citenamefont {Baye}\ and\ \citenamefont
  {Tursunov}(2011)}]{Baye2011}%
  \BibitemOpen
  \bibfield  {author} {\bibinfo {author} {\bibfnamefont {D.}~\bibnamefont
  {Baye}}\ and\ \bibinfo {author} {\bibfnamefont {E.}~\bibnamefont
  {Tursunov}},\ }\bibfield  {title} {\enquote {\bibinfo {title} {Beta delayed
  emission of a proton by a one-neutron halo nucleus},}\ }\href {\doibase
  10.1016/j.physletb.2010.12.069} {\bibfield  {journal} {\bibinfo  {journal}
  {Phys. Lett. B}\ }\textbf {\bibinfo {volume} {696}},\ \bibinfo {pages} {464
  -- 467} (\bibinfo {year} {2011})}\BibitemShut {NoStop}%
\bibitem [{\citenamefont {Riisager}\ \emph {et~al.}(2014)\citenamefont
  {Riisager} \emph {et~al.}}]{Riisager2014}%
  \BibitemOpen
  \bibfield  {author} {\bibinfo {author} {\bibfnamefont {K.}~\bibnamefont
  {Riisager}} \emph {et~al.},\ }\bibfield  {title} {\enquote {\bibinfo {title}
  {$^{11}${Be}($\beta$p), a quasi-free neutron decay?}}\ }\href {\doibase
  10.1016/j.physletb.2014.03.062} {\bibfield  {journal} {\bibinfo  {journal}
  {Phys. Lett. B}\ }\textbf {\bibinfo {volume} {732}},\ \bibinfo {pages} {305
  -- 308} (\bibinfo {year} {2014})}\BibitemShut {NoStop}%
\bibitem [{\citenamefont {Ayyad}\ \emph {et~al.}(2019)\citenamefont {Ayyad}
  \emph {et~al.}}]{Ayyad2019}%
  \BibitemOpen
  \bibfield  {author} {\bibinfo {author} {\bibfnamefont {Y.}~\bibnamefont
  {Ayyad}} \emph {et~al.},\ }\bibfield  {title} {\enquote {\bibinfo {title}
  {Direct observation of proton emission in $^{11}\mathrm{Be}$},}\ }\href
  {\doibase 10.1103/PhysRevLett.123.082501} {\bibfield  {journal} {\bibinfo
  {journal} {Phys. Rev. Lett.}\ }\textbf {\bibinfo {volume} {123}},\ \bibinfo
  {pages} {082501} (\bibinfo {year} {2019})}\BibitemShut {NoStop}%
\bibitem [{\citenamefont {Riisager}\ \emph {et~al.}(2020)\citenamefont
  {Riisager} \emph {et~al.}}]{Riisager2020}%
  \BibitemOpen
  \bibfield  {author} {\bibinfo {author} {\bibfnamefont {K.}~\bibnamefont
  {Riisager}} \emph {et~al.},\ }\bibfield  {title} {\enquote {\bibinfo {title}
  {Search for beta-delayed proton emission from $^{11}${Be}},}\ }\href
  {\doibase 10.1140/epja/s10050-020-00110-2} {\bibfield  {journal} {\bibinfo
  {journal} {Eur. Phys. J. A}\ }\textbf {\bibinfo {volume} {56}},\ \bibinfo
  {pages} {100} (\bibinfo {year} {2020})}\BibitemShut {NoStop}%
\bibitem [{NND()}]{NNDC}%
  \BibitemOpen
  \href@noop {} {}\bibinfo {note} {National Nuclear Data Center,
  \url{http://www.nndc.bnl.gov/}}\BibitemShut {NoStop}%
\bibitem [{\citenamefont {Elkamhawy}\ \emph {et~al.}(2021)\citenamefont
  {Elkamhawy}, \citenamefont {Yang}, \citenamefont {Hammer},\ and\
  \citenamefont {Platter}}]{Elkamhawy2019}%
  \BibitemOpen
  \bibfield  {author} {\bibinfo {author} {\bibfnamefont {W.}~\bibnamefont
  {Elkamhawy}}, \bibinfo {author} {\bibfnamefont {Z.}~\bibnamefont {Yang}},
  \bibinfo {author} {\bibfnamefont {H.-W.}\ \bibnamefont {Hammer}}, \ and\
  \bibinfo {author} {\bibfnamefont {L.}~\bibnamefont {Platter}},\ }\bibfield
  {title} {\enquote {\bibinfo {title} {$\beta$-delayed proton emission from
  $^{11}${Be} in effective field theory},}\ }\href {\doibase
  10.1016/j.physletb.2021.136610} {\bibfield  {journal} {\bibinfo  {journal}
  {Phys. Lett. B}\ }\textbf {\bibinfo {volume} {821}},\ \bibinfo {pages}
  {136610} (\bibinfo {year} {2021})}\BibitemShut {NoStop}%
\bibitem [{\citenamefont {Volya}(2020)}]{Volya2020}%
  \BibitemOpen
  \bibfield  {author} {\bibinfo {author} {\bibfnamefont {A.}~\bibnamefont
  {Volya}},\ }\bibfield  {title} {\enquote {\bibinfo {title} {Assessment of the
  beta-delayed proton decay rate of $^{11}${Be}},}\ }\href {\doibase
  10.1209/0295-5075/130/12001} {\bibfield  {journal} {\bibinfo  {journal}
  {Europhys. Lett.}\ }\textbf {\bibinfo {volume} {130}},\ \bibinfo {pages}
  {12001} (\bibinfo {year} {2020})}\BibitemShut {NoStop}%
\bibitem [{\citenamefont {Oko{\l}owicz}\ \emph {et~al.}(2020)\citenamefont
  {Oko{\l}owicz}, \citenamefont {P{\l}oszajczak},\ and\ \citenamefont
  {Nazarewicz}}]{Okolowicz2020}%
  \BibitemOpen
  \bibfield  {author} {\bibinfo {author} {\bibfnamefont {J.}~\bibnamefont
  {Oko{\l}owicz}}, \bibinfo {author} {\bibfnamefont {M.}~\bibnamefont
  {P{\l}oszajczak}}, \ and\ \bibinfo {author} {\bibfnamefont {W.}~\bibnamefont
  {Nazarewicz}},\ }\bibfield  {title} {\enquote {\bibinfo {title} {Convenient
  location of a near-threshold proton-emitting resonance in
  $^{11}\mathrm{B}$},}\ }\href {\doibase 10.1103/PhysRevLett.124.042502}
  {\bibfield  {journal} {\bibinfo  {journal} {Phys. Rev. Lett.}\ }\textbf
  {\bibinfo {volume} {124}},\ \bibinfo {pages} {042502} (\bibinfo {year}
  {2020})}\BibitemShut {NoStop}%
\bibitem [{\citenamefont {Refsgaard}\ \emph {et~al.}(2019)\citenamefont
  {Refsgaard}, \citenamefont {B{\"u}scher}, \citenamefont {Arokiaraj},
  \citenamefont {Fynbo}, \citenamefont {Raabe},\ and\ \citenamefont
  {Riisager}}]{Refsgaard2019}%
  \BibitemOpen
  \bibfield  {author} {\bibinfo {author} {\bibfnamefont {J.}~\bibnamefont
  {Refsgaard}}, \bibinfo {author} {\bibfnamefont {J.}~\bibnamefont
  {B{\"u}scher}}, \bibinfo {author} {\bibfnamefont {A.}~\bibnamefont
  {Arokiaraj}}, \bibinfo {author} {\bibfnamefont {H.~O.~U.}\ \bibnamefont
  {Fynbo}}, \bibinfo {author} {\bibfnamefont {R.}~\bibnamefont {Raabe}}, \ and\
  \bibinfo {author} {\bibfnamefont {K.}~\bibnamefont {Riisager}},\ }\bibfield
  {title} {\enquote {\bibinfo {title} {Clarification of large-strength
  transitions in the $\ensuremath{\beta}$ decay of $^{11}\mathrm{Be}$},}\
  }\href {\doibase 10.1103/PhysRevC.99.044316} {\bibfield  {journal} {\bibinfo
  {journal} {Phys. Rev. C}\ }\textbf {\bibinfo {volume} {99}},\ \bibinfo
  {pages} {044316} (\bibinfo {year} {2019})}\BibitemShut {NoStop}%
\bibitem [{\citenamefont {Utsuno}\ and\ \citenamefont {Chiba}(2011)}]{psdu}%
  \BibitemOpen
  \bibfield  {author} {\bibinfo {author} {\bibfnamefont {Y.}~\bibnamefont
  {Utsuno}}\ and\ \bibinfo {author} {\bibfnamefont {S.}~\bibnamefont {Chiba}},\
  }\bibfield  {title} {\enquote {\bibinfo {title} {Multiparticle-multihole
  states around $^{16}\mathrm{O}$ and correlation-energy effect on the shell
  gap},}\ }\href {\doibase 10.1103/PhysRevC.83.021301} {\bibfield  {journal}
  {\bibinfo  {journal} {Phys. Rev. C}\ }\textbf {\bibinfo {volume} {83}},\
  \bibinfo {pages} {021301} (\bibinfo {year} {2011})}\BibitemShut {NoStop}%
\bibitem [{\citenamefont {Lubna}\ \emph {et~al.}(2019)\citenamefont {Lubna},
  \citenamefont {Kravvaris}, \citenamefont {Tabor}, \citenamefont {Tripathi},
  \citenamefont {Volya}, \citenamefont {Rubino}, \citenamefont {Allmond},
  \citenamefont {Abromeit}, \citenamefont {Baby},\ and\ \citenamefont
  {Hensley}}]{fsu}%
  \BibitemOpen
  \bibfield  {author} {\bibinfo {author} {\bibfnamefont {R.~S.}\ \bibnamefont
  {Lubna}}, \bibinfo {author} {\bibfnamefont {K.}~\bibnamefont {Kravvaris}},
  \bibinfo {author} {\bibfnamefont {S.~L.}\ \bibnamefont {Tabor}}, \bibinfo
  {author} {\bibfnamefont {V.}~\bibnamefont {Tripathi}}, \bibinfo {author}
  {\bibfnamefont {A.}~\bibnamefont {Volya}}, \bibinfo {author} {\bibfnamefont
  {E.}~\bibnamefont {Rubino}}, \bibinfo {author} {\bibfnamefont {J.~M.}\
  \bibnamefont {Allmond}}, \bibinfo {author} {\bibfnamefont {B.}~\bibnamefont
  {Abromeit}}, \bibinfo {author} {\bibfnamefont {L.~T.}\ \bibnamefont {Baby}},
  \ and\ \bibinfo {author} {\bibfnamefont {T.~C.}\ \bibnamefont {Hensley}},\
  }\bibfield  {title} {\enquote {\bibinfo {title} {Structure of
  $^{38}\mathrm{Cl}$ and the quest for a comprehensive shell model
  interaction},}\ }\href {\doibase 10.1103/PhysRevC.100.034308} {\bibfield
  {journal} {\bibinfo  {journal} {Phys. Rev. C}\ }\textbf {\bibinfo {volume}
  {100}},\ \bibinfo {pages} {034308} (\bibinfo {year} {2019})}\BibitemShut
  {NoStop}%
\bibitem [{\citenamefont {Yuan}(2017)}]{Yuan2017}%
  \BibitemOpen
  \bibfield  {author} {\bibinfo {author} {\bibfnamefont {C.-X.}\ \bibnamefont
  {Yuan}},\ }\bibfield  {title} {\enquote {\bibinfo {title} {Impact of
  off-diagonal cross-shell interaction on $^{14}\mathrm{C}$},}\ }\href
  {\doibase 10.1088/1674-1137/41/10/104102} {\bibfield  {journal} {\bibinfo
  {journal} {Chin. Phys. C}\ }\textbf {\bibinfo {volume} {41}},\ \bibinfo
  {pages} {104102} (\bibinfo {year} {2017})}\BibitemShut {NoStop}%
\bibitem [{\citenamefont {Luo}\ \emph {et~al.}(2002)\citenamefont {Luo},
  \citenamefont {Okołowicz}, \citenamefont {Płoszajczak},\ and\ \citenamefont
  {Michel}}]{Luo2002}%
  \BibitemOpen
  \bibfield  {author} {\bibinfo {author} {\bibfnamefont {Y.}~\bibnamefont
  {Luo}}, \bibinfo {author} {\bibfnamefont {J.}~\bibnamefont {Okołowicz}},
  \bibinfo {author} {\bibfnamefont {M.}~\bibnamefont {Płoszajczak}}, \ and\
  \bibinfo {author} {\bibfnamefont {N.}~\bibnamefont {Michel}},\ }\bibfield
  {title} {\enquote {\bibinfo {title} {Shell model embedded in the continuum
  for binding systematics in neutron-rich isotopes of oxygen and fluor},}\
  }\href@noop {} {\bibfield  {journal} {\bibinfo  {journal}
  {arXiv:nucl-th/0211068}\ } (\bibinfo {year} {2002})}\BibitemShut {NoStop}%
\bibitem [{\citenamefont {Michel}\ \emph {et~al.}(2004)\citenamefont {Michel},
  \citenamefont {Nazarewicz}, \citenamefont {Okołowicz}, \citenamefont
  {Płoszajczak},\ and\ \citenamefont {Rotureau}}]{Michel2004}%
  \BibitemOpen
  \bibfield  {author} {\bibinfo {author} {\bibfnamefont {N.}~\bibnamefont
  {Michel}}, \bibinfo {author} {\bibfnamefont {W.}~\bibnamefont {Nazarewicz}},
  \bibinfo {author} {\bibfnamefont {J.}~\bibnamefont {Okołowicz}}, \bibinfo
  {author} {\bibfnamefont {M.}~\bibnamefont {Płoszajczak}}, \ and\ \bibinfo
  {author} {\bibfnamefont {J.}~\bibnamefont {Rotureau}},\ }\bibfield  {title}
  {\enquote {\bibinfo {title} {Shell model description of nuclei far from
  stability},}\ }\href@noop {} {\bibfield  {journal} {\bibinfo  {journal} {Acta
  Phys. Pol. B}\ }\textbf {\bibinfo {volume} {35}},\ \bibinfo {pages}
  {1249--1261} (\bibinfo {year} {2004})}\BibitemShut {NoStop}%
\bibitem [{\citenamefont {Charity}\ \emph {et~al.}(2018)\citenamefont {Charity}
  \emph {et~al.}}]{Charity2018}%
  \BibitemOpen
  \bibfield  {author} {\bibinfo {author} {\bibfnamefont {R.~J.}\ \bibnamefont
  {Charity}} \emph {et~al.},\ }\bibfield  {title} {\enquote {\bibinfo {title}
  {Spin alignment following inelastic scattering of $^{17}\mathrm{Ne}$,
  lifetime of $^{16}\mathrm{F}$, and its constraint on the continuum coupling
  strength},}\ }\href {\doibase 10.1103/PhysRevC.97.054318} {\bibfield
  {journal} {\bibinfo  {journal} {Phys. Rev. C}\ }\textbf {\bibinfo {volume}
  {97}},\ \bibinfo {pages} {054318} (\bibinfo {year} {2018})}\BibitemShut
  {NoStop}%
\bibitem [{\citenamefont {Calci}\ \emph {et~al.}(2016)\citenamefont {Calci},
  \citenamefont {Navr\'atil}, \citenamefont {Roth}, \citenamefont
  {Dohet-Eraly}, \citenamefont {Quaglioni},\ and\ \citenamefont
  {Hupin}}]{Calci2016}%
  \BibitemOpen
  \bibfield  {author} {\bibinfo {author} {\bibfnamefont {A.}~\bibnamefont
  {Calci}}, \bibinfo {author} {\bibfnamefont {P.}~\bibnamefont {Navr\'atil}},
  \bibinfo {author} {\bibfnamefont {R.}~\bibnamefont {Roth}}, \bibinfo {author}
  {\bibfnamefont {J.}~\bibnamefont {Dohet-Eraly}}, \bibinfo {author}
  {\bibfnamefont {S.}~\bibnamefont {Quaglioni}}, \ and\ \bibinfo {author}
  {\bibfnamefont {G.}~\bibnamefont {Hupin}},\ }\bibfield  {title} {\enquote
  {\bibinfo {title} {Can ab initio theory explain the phenomenon of parity
  inversion in $^{11}\mathrm{Be}$?}}\ }\href {\doibase
  10.1103/PhysRevLett.117.242501} {\bibfield  {journal} {\bibinfo  {journal}
  {Phys. Rev. Lett.}\ }\textbf {\bibinfo {volume} {117}},\ \bibinfo {pages}
  {242501} (\bibinfo {year} {2016})}\BibitemShut {NoStop}%
\bibitem [{\citenamefont {Choudhary}\ \emph {et~al.}(2020)\citenamefont
  {Choudhary}, \citenamefont {Srivastava},\ and\ \citenamefont
  {Navr\'atil}}]{Choudhary2020}%
  \BibitemOpen
  \bibfield  {author} {\bibinfo {author} {\bibfnamefont {P.}~\bibnamefont
  {Choudhary}}, \bibinfo {author} {\bibfnamefont {P.~C.}\ \bibnamefont
  {Srivastava}}, \ and\ \bibinfo {author} {\bibfnamefont {P.}~\bibnamefont
  {Navr\'atil}},\ }\bibfield  {title} {\enquote {\bibinfo {title} {Ab initio
  no-core shell model study of $^{10--14}\mathrm{B}$ isotopes with realistic
  $nn$ interactions},}\ }\href {\doibase 10.1103/PhysRevC.102.044309}
  {\bibfield  {journal} {\bibinfo  {journal} {Phys. Rev. C}\ }\textbf {\bibinfo
  {volume} {102}},\ \bibinfo {pages} {044309} (\bibinfo {year}
  {2020})}\BibitemShut {NoStop}%
\bibitem [{\citenamefont {Som\`a}\ \emph {et~al.}(2020)\citenamefont {Som\`a},
  \citenamefont {Navr\'atil}, \citenamefont {Raimondi}, \citenamefont
  {Barbieri},\ and\ \citenamefont {Duguet}}]{Soma2020}%
  \BibitemOpen
  \bibfield  {author} {\bibinfo {author} {\bibfnamefont {V.}~\bibnamefont
  {Som\`a}}, \bibinfo {author} {\bibfnamefont {P.}~\bibnamefont {Navr\'atil}},
  \bibinfo {author} {\bibfnamefont {F.}~\bibnamefont {Raimondi}}, \bibinfo
  {author} {\bibfnamefont {C.}~\bibnamefont {Barbieri}}, \ and\ \bibinfo
  {author} {\bibfnamefont {T.}~\bibnamefont {Duguet}},\ }\bibfield  {title}
  {\enquote {\bibinfo {title} {Novel chiral hamiltonian and observables in
  light and medium-mass nuclei},}\ }\href {\doibase
  10.1103/PhysRevC.101.014318} {\bibfield  {journal} {\bibinfo  {journal}
  {Phys. Rev. C}\ }\textbf {\bibinfo {volume} {101}},\ \bibinfo {pages}
  {014318} (\bibinfo {year} {2020})}\BibitemShut {NoStop}%
\bibitem [{\citenamefont {Fortune}(2012)}]{Fortune2012}%
  \BibitemOpen
  \bibfield  {author} {\bibinfo {author} {\bibfnamefont {H.~T.}\ \bibnamefont
  {Fortune}},\ }\bibfield  {title} {\enquote {\bibinfo {title} {Properties of
  the lowest 1/2${}^{+}$, {$T=3/2$} states in {$A=11$} nuclei},}\ }\href
  {\doibase 10.1103/PhysRevC.85.044304} {\bibfield  {journal} {\bibinfo
  {journal} {Phys. Rev. C}\ }\textbf {\bibinfo {volume} {85}},\ \bibinfo
  {pages} {044304} (\bibinfo {year} {2012})}\BibitemShut {NoStop}%
\bibitem [{\citenamefont {Lane}\ and\ \citenamefont {Thomas}(1958)}]{Lane1958}%
  \BibitemOpen
  \bibfield  {author} {\bibinfo {author} {\bibfnamefont {A.~M.}\ \bibnamefont
  {Lane}}\ and\ \bibinfo {author} {\bibfnamefont {R.~G.}\ \bibnamefont
  {Thomas}},\ }\bibfield  {title} {\enquote {\bibinfo {title} {R-matrix theory
  of nuclear reactions},}\ }\href {\doibase 10.1103/RevModPhys.30.257}
  {\bibfield  {journal} {\bibinfo  {journal} {Rev. Mod. Phys.}\ }\textbf
  {\bibinfo {volume} {30}},\ \bibinfo {pages} {257--353} (\bibinfo {year}
  {1958})}\BibitemShut {NoStop}%
\bibitem [{\citenamefont {Wilkinson}(1962)}]{Wilkinson1962}%
  \BibitemOpen
  \bibfield  {author} {\bibinfo {author} {\bibfnamefont {D.~H.}\ \bibnamefont
  {Wilkinson}},\ }\bibfield  {title} {\enquote {\bibinfo {title}
  {Self-consistent determination of the effective radii of heavy nuclei for
  alpha-particle emission},}\ }\href {\doibase 10.1103/PhysRev.126.648}
  {\bibfield  {journal} {\bibinfo  {journal} {Phys. Rev.}\ }\textbf {\bibinfo
  {volume} {126}},\ \bibinfo {pages} {648--653} (\bibinfo {year}
  {1962})}\BibitemShut {NoStop}%
\bibitem [{\citenamefont {Ayyad}\ \emph {et~al.}(2022)\citenamefont {Ayyad}
  \emph {et~al.}}]{PhysRevLett.129.012501}%
  \BibitemOpen
  \bibfield  {author} {\bibinfo {author} {\bibfnamefont {Y.}~\bibnamefont
  {Ayyad}} \emph {et~al.},\ }\bibfield  {title} {\enquote {\bibinfo {title}
  {Evidence of a near-threshold resonance in $^{11}\mathrm{B}$ relevant to the
  $\ensuremath{\beta}$-delayed proton emission of $^{11}\mathrm{Be}$},}\ }\href
  {\doibase 10.1103/PhysRevLett.129.012501} {\bibfield  {journal} {\bibinfo
  {journal} {Phys. Rev. Lett.}\ }\textbf {\bibinfo {volume} {129}},\ \bibinfo
  {pages} {012501} (\bibinfo {year} {2022})}\BibitemShut {NoStop}%
\bibitem [{\citenamefont {Lopez-Saavedra}\ \emph {et~al.}(2022)\citenamefont
  {Lopez-Saavedra} \emph {et~al.}}]{PhysRevLett.129.012502}%
  \BibitemOpen
  \bibfield  {author} {\bibinfo {author} {\bibfnamefont {E.}~\bibnamefont
  {Lopez-Saavedra}} \emph {et~al.},\ }\bibfield  {title} {\enquote {\bibinfo
  {title} {Observation of a near-threshold proton resonance in
  $^{11}\mathrm{B}$},}\ }\href {\doibase 10.1103/PhysRevLett.129.012502}
  {\bibfield  {journal} {\bibinfo  {journal} {Phys. Rev. Lett.}\ }\textbf
  {\bibinfo {volume} {129}},\ \bibinfo {pages} {012502} (\bibinfo {year}
  {2022})}\BibitemShut {NoStop}%
\end{thebibliography}%

\end{document}